\documentclass[twocolumn,pra]{revtex4}
\usepackage{graphicx,amssymb,amsbsy}

	\def\<{\langle} 	\def\>{\rangle}

\def\al{\alpha}				
		\def\Del{\Delta}
\def\eps{\epsilon}

 \def\Bp{{\mathbf p}}  
	
  \def\B0{{\mathbf 0}}

\def\be{\begin{equation}}	\def\ee{\end{equation}}
\def\bea{\begin{eqnarray}}	\def\eea{\end{eqnarray}}

\begin{document}
\title{Comment on ``Superfluidity in the interior-gap states''}
\author{W. Vincent Liu and Frank Wilczek}
\affiliation{Center for Theoretical Physics, Department of Physics,
Massachusetts Institute of Technology,
Cambridge, Massachusetts 02139}



\maketitle


Recently we analyzed ~\cite{Liu-Wilczek:03} the possiblity of a new,
hybrid superfluid-normal state of matter, the ``interior gap'' phase.
We studied idealized models that support the existence of this phase,
and also proposed that it might eventually be realized in cold atom
systems.  In a recent paper~\cite{Wu-Yip:03}, Wu and Yip presented
calculations that they interpreted as indicating an instability of the
interior gap phase.  Here we observe that their interpretation does
not apply to the situations of most physical interest.  Their
instability was derived under the implicit assumption that one should
perturb the relevant systems, which typically contain two distinct
species, at fixed values of the separate chemical potentials.  However
it can be, and generally is, appropriate to enforce different
constraints, involving fixed values of particle numbers.  Under these
conditions one often finds, as indicated for fixed particle numbers 
in~\cite{Liu-Wilczek:03} and for different but related conditions 
in~\cite{Shovkovy:03,GLW:03}, stable states of the kind we claimed.

We will be considering simple models with Hamiltonian 
\be 
H = \sum_{\Bp,\al=L,H} (\eps_{\Bp\al}-\mu_\al) a^\dag_{\Bp\al} a_{\Bp\al} +
\mbox{interaction}  \label{eq:Hnn}
\ee
where $\eps_{\Bp\al} = \Bp^2/2m_\al$ and the index $\al$  lables the 
light (L) and heavy (H) fermions, respectively.
For  a particle-conserving interaction, 
such that the theory supports a U(1)$\otimes$U(1) symmetry
transforming the phases of 
fermion operators $a_{\Bp L}$ and $a_{\Bp H}$ separately, 
the fermion number densities $n_{L,H}$ are independently conserved.   
The model Hamiltonian 
considered by Wu and Yip falls
into this class, with equality of chemical potentials $\mu_\al \equiv
p^2_{F\al}/2m_\al$ assumed.

To clarify the possibilities, let us distinguish 
three cases:

\paragraph{Both number densities fixed.}

The normal state can give way to a superfluid state, through a pairing
instability.  We showed in \cite{Liu-Wilczek:03} that once the
coupling stength is greater than a critical value (which approaches
zero in the limit $m_H \gg m_L$) pairing is favored.  When both
denisities are conserved, however, the density difference $\Del n=n_H-
n_L$ cannot be affected by pairing.  Therefore the chemical potentials
must be adjusted to accommodate the change of band structure due to
pairing, holding particle densities fixed.  For example, if $p_{F}^H >
p_{F}^L$ the new $p^H_{F}$ must be enlarged to accommodate the
particles promoted from the interior.  Formally, one finds that a
branch of the Bogoliubov excitation spectrum must crosses zero energy,
and these zero-crossings determine the effective ``Fermi'' surfaces.
Whether the system chooses the BCS or interior gap state is really
determined by the initial values of particle densities.

In Ref.~\cite{Wu-Yip:03}, Wu and Yip assumed that the Fermi momenta
$p_{F\al}$ for both heavy and light fermions are unchanged across the
phase transition from the normal state to the superfluid state (being
either BCS or interior gap) or in the presence of a superflow current.
This is an inappropriate assumption for ultracold atoms confined in a trap
with negligible particle loss.  For example, if they
had been fixing  both heavy and light particle
densities, or the overall particle density (see next paragraph),  
their expression for the 
paramagnetic
current in linear response (c.f. eq.~(12) of Ref.~\cite{Wu-Yip:03}) would
employ modified chemical potentials in the presence of superflow
current,
\be
{1\over m_\al} \sum_\Bp \Bp \left[\tilde{\phi}^2_\al f(\eta_\al
  \tilde{E}_h +\Bp\cdot \mathbf{w}) -
   \phi^2_\al f(\eta_\al {E}_h) \right] \,,
\ee
with the tilde `$\sim$' indicating the adjusted $\tilde{p}_{F\al}$'s. 
The $\tilde{p}_{F\al}$'s receive
corrections in linear order $\sim |\mathbf{w}|$~\cite{GLW:03}, and so does the
paramagnetic current.  This is of the same order as, and counters, the
linear term arising directly  from $\Bp\cdot \mathbf{w}$.

\paragraph{Overall density fixed.} 

If we add a term proportional to
\be
a^\dag_{\Bp H} a_{\Bp L} +h.c. \,,
\ee
as for instance to describe Rabi oscillations between different
hyperfine spin
states of an alkali atom, 
then the symmetry of the Hamiltonian (\ref{eq:Hnn}) is reduced
to a single U(1), conjugate to the total density $n_t$. The relative
density $\Del n=n_L-n_H$ is no longer conserved.  Again, the interior
gap phase can be favorable in weak coupling.  A specific proposal is spelled out in ~\cite{LWZ:03}.

When pairing occurs, the chemical potential $\bar{\mu} =(\mu_L+\mu_H)/2$
must adjust to accommodate the total number density.  In a practical
experimental arrangement, the relative chemical potential $\Del\mu
\equiv \mu_L -\mu_H$ can be held fixed \cite{LWZ:03}. Then the
thermodynamic variables are $(n_t, \Del \mu)$.  $\Del p_F$, defined as
Fermi momentum difference, changes its value in response to pairing
when $m_L\neq m_H$, even if $\Del\mu$ is held fixed. 

\paragraph{Chemical potentials fixed.} 

If the model system (\ref{eq:Hnn}) is connected to reservoirs of both
light and heavy particles, then the particle numbers of both species
are not conserved within the subsystem.  We have realized for some
time that the interior gap state is very unlikely to occur in the weak
coupling regime for this case.  A detailed discussion is contained in
Ref.~\cite{GLW:03}. 

In Ref.~\cite{Liu-Wilczek:03} our discussion was mainly directed
toward cold atom systems with different kinds of atoms, which fall
under the first case mentioned above.  We also briefly mentioned the
possibility of interior gap state connecting two bands in an
electronic solid material.  That falls under the second case, since at
some level only overall electron number is strictly conserved.  The
third case, of which Wu and Yip analyzed a special example, did not
arise.

We are grateful for discussions with J. Bowers, E. Gubankova,
K. Rajagopal, and P. Zoller. 
This work is supported in part by funds provided by the
U.S. Department of Energy (D.O.E.) under cooperative research
agreement \#DF-FC02-94ER40818.


\bibliographystyle{apsrev} 
\bibliography{comment_yip3}

\begin{thebibliography}{5}
\expandafter\ifx\csname natexlab\endcsname\relax\def\natexlab#1{#1}\fi
\expandafter\ifx\csname bibnamefont\endcsname\relax
  \def\bibnamefont#1{#1}\fi
\expandafter\ifx\csname bibfnamefont\endcsname\relax
  \def\bibfnamefont#1{#1}\fi
\expandafter\ifx\csname citenamefont\endcsname\relax
  \def\citenamefont#1{#1}\fi
\expandafter\ifx\csname url\endcsname\relax
  \def\url#1{\texttt{#1}}\fi
\expandafter\ifx\csname urlprefix\endcsname\relax\def\urlprefix{URL }\fi
\providecommand{\bibinfo}[2]{#2}
\providecommand{\eprint}[2][]{\url{#2}}

\bibitem[{\citenamefont{Liu and Wilczek}(2003)}]{Liu-Wilczek:03}
\bibinfo{author}{\bibfnamefont{W.~V.} \bibnamefont{Liu}} \bibnamefont{and}
  \bibinfo{author}{\bibfnamefont{F.}~\bibnamefont{Wilczek}},
  \bibinfo{journal}{Phys. Rev. Lett.} \textbf{\bibinfo{volume}{90}},
  \bibinfo{pages}{047002} (\bibinfo{year}{2003}).

\bibitem[{Wu-()}]{Wu-Yip:03}
\bibinfo{note}{S.T. Wu and S. Yip, Phys. Rev. A (to appear) [arXiv:
  cond-mat/0303185].}

\bibitem[{Sho()}]{Shovkovy:03}
\bibinfo{note}{I. Shovkovy and M. Huang, arXiv: hep-ph/0302142.}

\bibitem[{\citenamefont{Gubankova et~al.}()\citenamefont{Gubankova, Liu, and
  Wilczek}}]{GLW:03}
\bibinfo{author}{\bibfnamefont{E.}~\bibnamefont{Gubankova}},
  \bibinfo{author}{\bibfnamefont{W.~V.} \bibnamefont{Liu}}, \bibnamefont{and}
  \bibinfo{author}{\bibfnamefont{F.}~\bibnamefont{Wilczek}},
  \bibinfo{note}{arXiv: hep-ph/0304016}.

\bibitem[{LWZ()}]{LWZ:03}
\bibinfo{note}{W. V. Liu, F. Wilczek, and P. Zoller (to appear).}

\end{thebibliography}

\end{document}